\documentclass{aaedp}
\usepackage{graphicx}
\begin{document}
   \title{The Dynamical Status of the Cluster of Galaxies 
1E0657-56\thanks{Based on observations collected at the European
Southern Observatory (La Silla, Chile), Proposal ID: 64.0-0664}}

   \author{R. Barrena\inst{1} 
          \and
          A. Biviano\inst{2}
          \and
          M. Ramella\inst{2}
          \and
          E.E. Falco\inst{3}
          \and
          S. Seitz\inst{4}
         }

   \offprints{A. Biviano, \email{abiviano@ts.astro.it}}

   \institute{Instituto de Astrof\'{\i}sica de Canarias, E-38200 La Laguna,
              Tenerife, Spain\\
         \and
             INAF, Osservatorio Astronomico di Trieste, via G.B. Tiepolo 11, 
	      I-34131 Trieste, Italy\\
         \and
             Smithsonian Institution, F.L. Whipple Observatory, P.O. Box97,
	      670 Mount Hopkins, Amado, AZ 85645\\
         \and
             Universit\"ats-Sternwarte, Scheinerstrasse 1, D-81679 Muenchen, 
              Germany\\
             }

   \date{Received / Accepted }

   \abstract{We present the results of a new spectroscopic and
             photometric survey of the hot X-ray cluster
             \mbox{1E0657-56}, at $z=0.296$. We determine the presence
             of a low velocity dispersion subcluster, which is offset
             from the main cluster position by 0.7 Mpc and $\simeq 600$
             km~s$^{-1}$. We determine the virial masses and total
             luminosities of the cluster and its subcluster, and solve
             for the two-body dynamical model. With additional
             constraints from the results of the analysis of the
             cluster X-ray emission by Markevitch et al.
             (\cite{mar01}), we find that the subcluster passed
             through the cluster centre $\simeq 0.15$ Gyr ago. Taken
             at face value the mass of the subcluster is typical of a
             loose group. It is however difficult to establish the
             pre-merger mass of the colliding system. We provide
             tentative evidence that the subcluster is in fact the
             remnant core of a moderately massive cluster, stripped by
             the collision with \mbox{1E0657-56}. The main cluster
             dynamics does not seem to have suffered from this
             collision. On the contrary, the cluster X-ray properties
             seem to have been significantly affected.  We also
             discuss the effect of the subcluster collision in
             relation to starburst events and the cluster radio halo.
              }

   \maketitle
%

\section{Introduction}
The evolution of clusters of galaxies as seen in numerical simulations
is characterized by the asymmetric accretion of mass clumps from
surrounding filaments (e.g. Diaferio et al. \cite{dia01}). Nearby
clusters are characterized by a variety of morphologies, indicative of
different dynamical properties. Distant clusters, at
redshifts $z>0.8$, are often characterized by an elongated
distribution, traced by several, apparently distinct, galaxy
clumps. Such is the case of Cl~0023+0423 at $z=0.84$ (Lubin et
al. \cite{lub98}), RX~J1716.6+6708 at $z=0.81$ (Gioia et
al. \cite{gio99}), MS~1054-03 at $z=0.84$ (van Dokkum et al.
\cite{vdo00}). Other distant clusters are found to have nearby
companions, possibly in a pre-merger phase (Lubin et al. \cite{lub00};
Rosati et al. \cite{ros99}; Haines et al. \cite{hai01}; Pentericci et
al. \cite{pen00}). Most of these high-$z$ clusters are X-ray selected,
and are therefore expected to be very massive.

1E0657-558 is in many respects a low-redshift ($z=0.296$) analogue of
these high-$z$ clusters. It is X-ray selected, it has an elongated
morphology, and there is evidence for an additional subclump located
to the West with respect to the main cluster region (see
Sect.~\ref{galdis}).  Its high X-ray luminosity and temperature
(Tucker et al. \cite{tuc98}, hereafter T98; Liang et al.
\cite{lia00}; Markevitch et al. \cite{mark01}, hereafter M01) as well
as its high velocity dispersion (T98) strongly suggest it to be a very
massive cluster. A detailed dynamical study of 1E0657-558 could
therefore help us understand the dynamics of its more distant
analogues.

1E0657-558 is also very interesting {\em per se.} After detection in
X-ray by the {\em Einstein} IPC, {\em ROSAT} and {\em ASCA}
observations constrained its X-ray temperature to be $kT=17.4 \pm 2.5$
keV (T98) or slightly lower -- $kT=14.5^{+2.0}_{-1.7}$ keV, according
to Liang et al. (\cite{lia00}). \mbox{1E0657-56} is thus one of the
hottest clusters known.  It displays an irregular X-ray morphology,
with two major emission peaks, both clearly offset from the main
overdensity of projected galaxy counts (see Fig.5 in Liang et
al. \cite{lia00} and Fig.\ref{rad_x} in this paper). Recently, M01
have reported on {\em Chandra} observations of \mbox{1E0657-56}. They
confirm the high temperature of this cluster ($kT=14.8^{+1.7}_{-1.2}$
keV), and provide evidence for a compact subcluster at lower
temperature ($kT \sim 6$--7 keV).  According to M01, this subcluster
is seen 0.1--0.2 Gyr after its collision with the main cluster core.

The radio halo of \mbox{1E0657-56} was recently detected by Liang et
al. (\cite{lia00}). The most widely accepted scenario for radio halo
production requires the acceleration of thermal electrons to
ultra-relativistic energies, and amplification of the intra-cluster
magnetic field, by an energetic cluster-cluster collision (e.g. Liang
\cite{lia00b}). It is important to establish if the radio halo of
\mbox{1E0657-56} is in any way related to the recent collision with the
X-ray subcluster identified by M01.

In the optical, 1E0657-558 was detected by Tucker et
al. (\cite{tuc95}), who also revealed a luminous giant gravitational
arc, an additional evidence for a strong mass concentration. The arc
was confirmed by follow-up observations at the ESO {\em New Technology
Telescope} (NTT), that also provided a first tentative estimate of the
cluster velocity dispersion, $\sigma_v=1213^{+352}_{-191}$ km~s$^{-1}$
(T98). Follow-up observations with FORS@VLT were obtained at the end
of 1998, providing the spectrum and redshift of the giant arc
\footnote{See http://www.eso.org/outreach/press-rel/pr-1999/phot-16-99.html.}.

In January 2000 we obtained additional spectra of galaxies in the
cluster region, with the purpose of constraining its dynamical
status. To this end, in this paper we present an analysis of the
phase-space distribution of cluster member galaxies. We also consider
the relative frequencies and distributions of galaxies of different
morphological and spectral types, which are useful indicators of a
cluster's dynamical status (e.g. Moss \& Whittle \cite{mos00}).  Along
with the additional information from the X-ray and radio observations,
our new spectroscopic observations allow us to posit a plausible
scenario for the dynamical status of 1E0657-558.

The plan of this paper is the following. In Sect.~\ref{obser} we
describe our new photometric and spectroscopic observations. In
Sect.~\ref{anares} we provide our results: a) we determine the
distributions of cluster members in the spatial coordinates and in
velocities, and provide evidence for the existence of a subcluster; b)
we also determine the masses, luminosities, and mass-to-light ratios
of the main cluster and its subcluster, and solve the two-body problem
for these systems; c) we then consider the relative frequencies and
distributions of cluster members of different morphologies and
spectral types. In Sect.~\ref{discuss} we provide our interpretation
of these results, also taking into account the results from the X-ray
and radio observations. We summarize our results in Sect.~\ref{summa}.

Throughout this paper, we use $H_0=70$ km s$^{-1}$ Mpc$^{-1}$
(e.g. Freedman et al. \cite{fre01}; Liu \& Graham \cite{liu01}) in a
flat cosmology with $\Omega_0=0.3$ and $\Omega_{\Lambda}=0.7$
(e.g. Bahcall et al. \cite{bah99}). In the adopted cosmology, 1~arcmin
corresponds to 0.26~Mpc at the cluster distance.

\section{Data}
\label{obser}
We carried out the spectroscopic observations at the ESO NTT in La
Silla, during two nights in January 2000.  The weather conditions were
good, with seeing slightly below $1''$.  We observed with the red-arm
of EMMI in Multi-Object Spectroscopy (MOS) mode.  EMMI was equipped
with a Tektronix TK2048 CCD of $2048 \times 2047$ 24 $\mu$m pixels,
allowing for an unvignetted field-of-view of $5' \times 8.6'$.  We
used grism \#2, giving a wavelength coverage from 3850 \AA ~to 9000
\AA, and a dispersion of 2.8 \AA/pixel.

We took spectra for 129 targets with 5 MOS masks, with exposure times
between 2700 and 7200 sec per mask. We reduced the data with standard
IRAF\footnote{IRAF is distributed by the National Optical Astronomy
Observatories, which is operated by AURA Inc. under contract with
NSF.}  packages. All spectra were also visually examined, to
exclude possible misidentification of night-sky lines and residuals
from cosmic-ray impacts with real spectral features.  The
signal-to-noise ratios of our spectra range from 8 to 27, with an
average S/N$\sim$17. We determined redshifts for 104 galaxies, using
the IRAF tasks XCSAO and EMSAO. Our redshifts span the range
0.0484--0.4827 with an average error of 0.0003. We also add the
redshifts of the 16 galaxies that were observed in December 1993 with
the same instrumentation, but a slightly different set-up, resulting
in a lower resolution (5.9 \AA/pixel). These 16 galaxy redshifts were
already used by T98 to compute the velocity dispersion of the cluster.

We also determined the equivalent widths (EW hereafter), or upper
limits, of the absorption line H$\delta$ and the emission line
[\ion{O}{ii}], in order to classify post-starburst and starburst
galaxies (see Sect.~\ref{morph}). We estimated the minimum measurable EW of
each spectrum as the width of a line spanning 2.8 \AA ~(our
dispersion) in wavelength, with an intensity three times the rms
noise in the adjacent continuum.

\begin{table*}
\centering
\caption[]{Data for 1E0657-056 cluster members}
\label{data}
\begin{tabular}{ccccccccrrc}
\hline
\hline
 ID  & $\alpha_{\rm{J2000}}$ & $\delta_{\rm{J2000}}$ & $B$ & $B-R$ & $B-I$ & $cz_{\odot}$ 
& $\delta_{\rm{cz}}$ & EW(H$\delta$) & EW([\ion{O}{ii}]) & Type \\
 & hh:mm:ss & $^{\circ} \ ' \ ''$ & & & & km s$^{-1}$ & km s$^{-1}$ & \AA & \AA & (Early/Late) \\
\hline
01 & 06:57:56 & -55:54:24 &   --  &  --  &  --  & 87495 &  38 & $<3.0$ &   --   & --\\
02 & 06:57:56 & -55:55:55 &   --  &  --  &  --  & 88185 &  38 & $<2.6$ &   --   & --\\
03 & 06:57:57 & -55:54:37 &   --  &  --  &  --  & 88431 &  54 & $<2.1$ &   --   & --\\
04 & 06:58:00 & -55:53:40 &   --  &  --  &  --  & 86311 &  41 &    3.6 &   --   & --\\
05 & 06:58:08 & -55:53:36 &   --  &  --  &  --  & 91966 &  52 &    5.3 &   --   & --\\
06 & 06:58:08 & -55:55:34 & 22.97 & 3.02 & 3.16 & 88245 &  43 & $<2.3$ &   --   & E \\
07 & 06:58:12 & -55:54:30 & 22.36 & 2.54 & 2.97 & 89991 &  33 &    6.3 &   --   & E \\
08 & 06:58:14 & -55:56:37 & 21.61 & 2.75 & 3.92 & 89383 &  48 & $<2.6$ &   --   & E \\
09 & 06:58:16 & -55:56:01 & 22.96 & 2.65 & 3.82 & 89215 &  52 & $<2.8$ &   --   & E \\
10 & 06:58:16 & -55:56:37 & 20.43 & 2.48 & 3.84 & 89172 &  50 & $<2.0$ &   --   & E \\
11 & 06:58:18 & -55:54:59 & 21.88 & 3.01 & 3.55 & 87957 &  44 & $<1.6$ &   --   & E \\
12 & 06:58:18 & -55:56:15 & 22.68 & 2.79 & 3.95 & 89865 &  53 & $<2.6$ &   --   & E \\
13 & 06:58:18 & -55:56:36 & 21.81 & 2.82 & 4.04 & 89458 &  46 & $<2.0$ &   --   & E \\
14 & 06:58:19 & -55:55:52 & 22.25 & 2.32 & 3.40 & 87846 &  34 & $<2.0$ &   --   & E \\
15 & 06:58:20 & -55:54:59 & 21.89 & 2.26 & 2.83 & 88689 &  52 & $<3.3$ &   --   & E \\
16 & 06:58:20 & -55:55:56 & 22.63 & 2.50 & 3.54 & 89798 &  44 & $<2.6$ &   --   & E \\
17 & 06:58:20 & -55:56:32 & 22.79 & 2.61 & 3.76 & 89514 &  61 & $<2.6$ &   --   & E \\
18 & 06:58:21 & -55:54:47 & 21.80 & 2.00 & 1.84 & 90735 &  34 & $<2.1$ &   --   & E \\
19 & 06:58:22 & -55:57:12 & 21.40 & 2.44 & 3.60 & 87360 &  36 & $<2.4$ &   --   & L \\
20 & 06:58:23 & -55:54:55 & 21.72 & 2.24 & 3.30 & 85411 &  29 &    4.9 &   --   & E \\
21 & 06:58:24 & -55:56:29 & 22.36 & 2.40 & 3.56 & 90338 &  58 & $<3.3$ &   --   & E \\
22 & 06:58:25 & -55:56:51 & 21.79 & 2.52 & 3.70 & 88106 &  40 &    4.6 &   --   & L \\
23 & 06:58:28 & -55:55:46 & 22.25 & 2.36 & 3.59 & 85601 &  51 &    8.8 &   --   & E \\
24 & 06:58:29 & -55:55:57 & 22.08 & 2.19 & 3.21 & 86082 &  57 &    6.4 &   --   & E \\
25 & 06:58:29 & -55:56:47 & 19.84 & 2.36 & 3.52 & 87585 &  35 & $<2.0$ &   --   & L \\
26 & 06:58:30 & -55:56:05 & 22.01 & 2.25 & 3.40 & 86349 &  38 & $<2.4$ &   --   & E \\
27 & 06:58:30 & -55:57:52 & 22.29 & 2.68 & 3.88 & 88729 &  53 & $<2.3$ &   --   & E \\
28 & 06:58:31 & -55:54:38 & 22.71 & 2.43 & 3.59 & 88691 &  50 & $<2.4$ &   --   & E \\
29 & 06:58:31 & -55:56:52 & 21.81 & 2.10 & 3.13 & 85593 &  50 & $<2.6$ &   --   & L \\
30 & 06:58:31 & -55:57:59 & 19.73 & 2.09 & 3.06 & 87239 &  46 &    4.8 &   --   & L \\
31 & 06:58:32 & -55:55:48 & 22.52 & 2.35 & 3.49 & 90096 &  52 &    5.9 &   --   & E \\
32 & 06:58:32 & -56:00:03 & 21.09 & 2.27 & 3.32 & 89103 &  64 &    7.0 &   --   & E \\
33 & 06:58:35 & -55:57:05 & 22.19 & 2.62 & 3.80 & 92134 &  44 & $<2.0$ &   --   & E \\
34 & 06:58:35 & -55:58:44 & 22.04 & 2.32 & 3.45 & 87638 &  44 &    3.8 &   --   & E \\
35 & 06:58:36 & -55:55:09 & 20.62 & 2.39 & 3.52 & 90517 &  38 & $<1.3$ &   --   & L \\
36 & 06:58:36 & -55:57:18 & 22.49 & 2.64 & 3.79 & 86925 &  38 & $<2.4$ &   --   & E \\
37 & 06:58:37 & -55:55:17 & 21.41 & 2.16 & 3.21 & 87693 &  42 &    6.3 &   --   & E \\
38 & 06:58:37 & -55:56:18 & 22.44 & 3.36 & 4.59 & 88872 &  29 & $<2.6$ &   --   & E \\
39 & 06:58:37 & -55:56:21 & 22.93 & 2.41 & 3.63 & 89349 &  67 & $<3.6$ &   --   & E \\
40 & 06:58:37 & -55:56:24 & 20.17 & 1.37 & 2.53 & 86954 &  27 &   12.3 &   --   & E \\
41 & 06:58:37 & -55:56:32 & 22.84 & 3.36 & 4.64 & 90698 &  39 & $<2.1$ &   --   & E \\
42 & 06:58:37 & -55:56:48 & 21.19 & 2.43 & 3.60 & 88320 &  45 & $<3.3$ &   --   & E \\
43 & 06:58:37 & -55:57:03 & 21.58 & 2.54 & 2.74 & 88706 &  23 & $<2.4$ &   --   & E \\
44 & 06:58:37 & -55:58:58 & 22.35 & 3.05 & 4.35 & 87461 &  39 &    7.8 &   --   & E \\
45 & 06:58:38 & -55:56:27 & 22.16 & 3.19 & 4.41 & 89546 &  35 &    3.4 &   --   & E \\
46 & 06:58:38 & -55:57:23 & 21.00 & 3.22 & 3.77 & 87403 &  28 & $<1.8$ &   --   & E \\
47 & 06:58:38 & -55:57:32 & 21.79 & 1.93 & 3.55 & 91580 &  99 & $<2.0$ &   --   & L \\
48 & 06:58:38 & -55:57:46 & 21.54 & 2.65 & 3.51 & 90765 & 104 & $<1.4$ &   --   & E \\
49 & 06:58:38 & -55:59:01 & 22.38 & 2.96 & 3.95 & 88909 & 102 &   13.6 &   --   & L \\
50 & 06:58:39 & -55:57:32 & 22.01 & 2.66 & 4.43 & 89266 &  37 & $<2.3$ &   --   & E \\
51 & 06:58:40 & -55:56:13 & 22.53 & 2.54 & 3.68 & 88818 &  32 & $<2.4$ &   --   & E \\
52 & 06:58:40 & -55:57:48 & 23.47 & 2.28 & 3.35 & 86092 &  77 & $<4.5$ &   --   & L \\
53 & 06:58:40 & -55:59:20 & 22.05 & 2.27 & 3.42 & 90912 &  64 & $<2.4$ &   --   & E \\
54 & 06:58:41 & -55:57:00 & 22.22 & 2.47 & 3.64 & 88240 &  37 & $<1.9$ &   --   & E \\
55 & 06:58:41 & -55:57:35 & 23.20 & 2.55 & 3.70 & 87485 &  52 & $<3.6$ &   --   & E \\
\hline
\end{tabular}
\end{table*}

\addtocounter{table}{-1}
\begin{table*}
\caption[ ]{Continued.}
\begin{tabular}{ccccccccrrc}
\hline
 ID  & $\alpha_{\rm{J2000}}$ & $\delta_{\rm{J2000}}$ & $B$ & $B-R$ & $B-I$ & $cz_{\odot}$ 
& $\delta_{\rm{cz}}$ & EW(H$\delta$) & EW([\ion{O}{ii}]) & Type \\
 & hh:mm:ss & $^{\circ} \ ' \ ''$ & & & & km s$^{-1}$ & km s$^{-1}$ & \AA & \AA & (Early/Late) \\
\hline
56 & 06:58:41 & -55:59:04 & 23.43 & 2.49 & 3.62 & 87713 &  65 & $<1.7$ &   --   & E \\ 
57 & 06:58:42 & -55:59:20 & 21.17 & 1.41 & 2.41 & 87300 &  10 & $<2.8$ &  26.3  & L \\ 
58 & 06:58:42 & -56:00:28 & 23.00 & 2.64 & 3.34 & 91988 &  82 &   13.2 &  27.0  & L \\  
59 & 06:58:43 & -55:58:36 & 19.92 & 2.62 & 3.82 & 89893 &  51 & $<1.9$ &   --   & E \\ 
60 & 06:58:44 & -55:57:22 & 22.49 & 2.22 & 3.33 & 89614 &  87 & $<2.0$ &   --   & E \\ 
61 & 06:58:45 & -55:57:57 & 22.21 & 2.64 & 3.68 & 87582 &  29 & $<1.6$ &   --   & E \\ 
62 & 06:58:45 & -55:58:35 & 22.26 & 3.70 & 4.92 & 90378 &  39 & $<1.9$ &   --   & E \\ 
63 & 06:58:45 & -55:59:42 & 22.53 & 2.33 & 3.65 & 89841 &  62 & $<2.8$ &   --   & E \\  
64 & 06:58:46 & -55:58:38 & 21.89 & 3.06 & 2.85 & 90785 &  63 & $<2.6$ &   --   & E \\ 
65 & 06:58:49 & -55:59:04 & 21.81 & 2.33 & 3.50 & 89252 &  41 & $<2.6$ &   --   & E \\  
66 & 06:58:51 & -56:00:25 & 22.49 & 3.26 & 4.78 & 89400 &  32 & $<1.9$ &   --   & E \\  
67 & 06:58:51 & -56:00:26 &   --  &  --  &  --  & 89293 &  42 & $<1.7$ &   --   & --\\ 
68 & 06:58:27 & -56:00:00 & 20.07 & 2.26 & 2.77 & 89928 & 231 &   --   &   --   & E \\
69 & 06:58:31 & -55:56:04 & 21.34 & 2.15 & 3.21 & 87874 & 212 &   --   &   --   & L \\
70 & 06:58:32 & -55:56:37 & 21.49 & 2.66 & 3.88 & 86876 & 163 &   --   &   --   & E \\
71 & 06:58:33 & -55:56:36 & 20.93 & 2.70 & 3.87 & 88584 & 159 &   --   &   --   & E \\
72 & 06:58:34 & -55:56:19 & 20.61 & 2.50 & 3.65 & 90184 & 176 &   --   &   --   & L \\
73 & 06:58:35 & -55:56:57 & 20.47 & 2.82 & 4.01 & 88897 & 163 &   --   &   --   & E \\
74 & 06:58:35 & -55:57:20 & 22.26 & 2.63 & 3.73 & 89956 & 276 &   --   &   --   & E \\
75 & 06:58:36 & -55:56:59 & 21.69 & 2.80 & 3.78 & 89430 & 220 &   --   &   --   & E \\
76 & 06:58:38 & -55:57:26 & 20.32 & 2.54 & 3.76 & 89016 & 201 &   --   &   --   & E \\
77 & 06:58:40 & -55:56:04 & 21.24 & 1.93 & 2.93 & 87753 & 193 &   --   &   --   & E \\
78 & 06:58:42 & -55:57:51 & 21.79 & 2.39 & 3.52 & 89388 & 138 &   --   &   --   & E \\
\hline                                                           
\end{tabular}
\end{table*}

We obtained exposures on a $6' \times 6'$ field
centered on $\alpha$=06:58:29 and $\delta=-$55:57:22 (J2000) using
FORS1 images through the $B$, $R$ and $I$ Bessel filters, at the ESO
VLT in December 1998.  The exposure times were 600 sec in each band;
the images reach the $B=24$, $R=23.5$, $I=21.5$ completeness
magnitudes. We carried out the photometric analysis using the {\it
SExtractor} package (Bertin \& Arnouts \cite{ber96}). We also
determined rough morphological types by visual inspection.

Our spectroscopic sample is 65 \% (45 \%) complete down to $I=19.0$
($I=20$), within an elongated region of 20 arcmin$^2$ around the
cluster main body.

In Table~\ref{data} we list the data for cluster members (membership
is defined in Sect.~\ref{galdis}). In Col. (1) we give an
identification number, in Cols. (2) and (3) the right ascension and
declination (J2000), in the next three Cols. we give the $B$
magnitudes and $B-R$, $B-I$ colours. In Cols. (7) and (8) we list
heliocentric velocities and their errors, respectively. In Cols. (9)
and (10) we list the EWs of H$\delta$ and [\ion{O}{ii}],
respectively. Finally, in Col. (11) we provide the galaxy
morphologies ('E' stands for 'early-type' and 'L' for
'late-type'). Data for the last 11 galaxies were taken from T98.

\section{Analysis \& Results}
\label{anares}
\subsection{Galaxy Distribution and Subclustering}
\label{galdis}
\begin{figure}
\centering
\includegraphics[width=9cm]{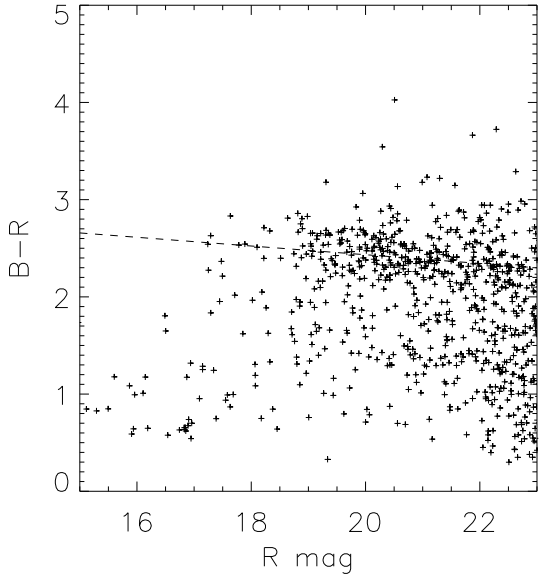}
\caption{$B-R$ vs $R$ distribution of the galaxies within the cluster 
field. The dashed line is a fit to the points with
$17<R<21.5$ and $2.1<B-R<2.8$, $B-R=3.33-0.045 \, R$, and defines the
CM sequence of the early-type population of cluster members.}
\label{BRR}
\end{figure}

We use the photometric sample to determine the projected spatial
distribution of cluster members. To this purpose, we select galaxies
in the colour-magnitude (CM) diagram, $B-R$ vs. $R$ (see
Fig.\ref{BRR}).  The CM diagram clearly shows the red sequence of
early-type galaxies in the cluster.  Taking into account the relative
distance modulus, evolutionary- and K-correction, we compare the $B-R$
vs. $R$ CM sequence of \mbox{1E0657-56} with that of Coma (Mazure et
al. \cite{maz88}).  We adopt Poggianti's (\cite{pog97}) evolutionary-
and K-correction for a passively evolving elliptical.  We find
excellent agreement between the CM sequence of \mbox{1E0657-56} and
that of Coma.

\begin{figure*}
\centering
\includegraphics[width=\textwidth]{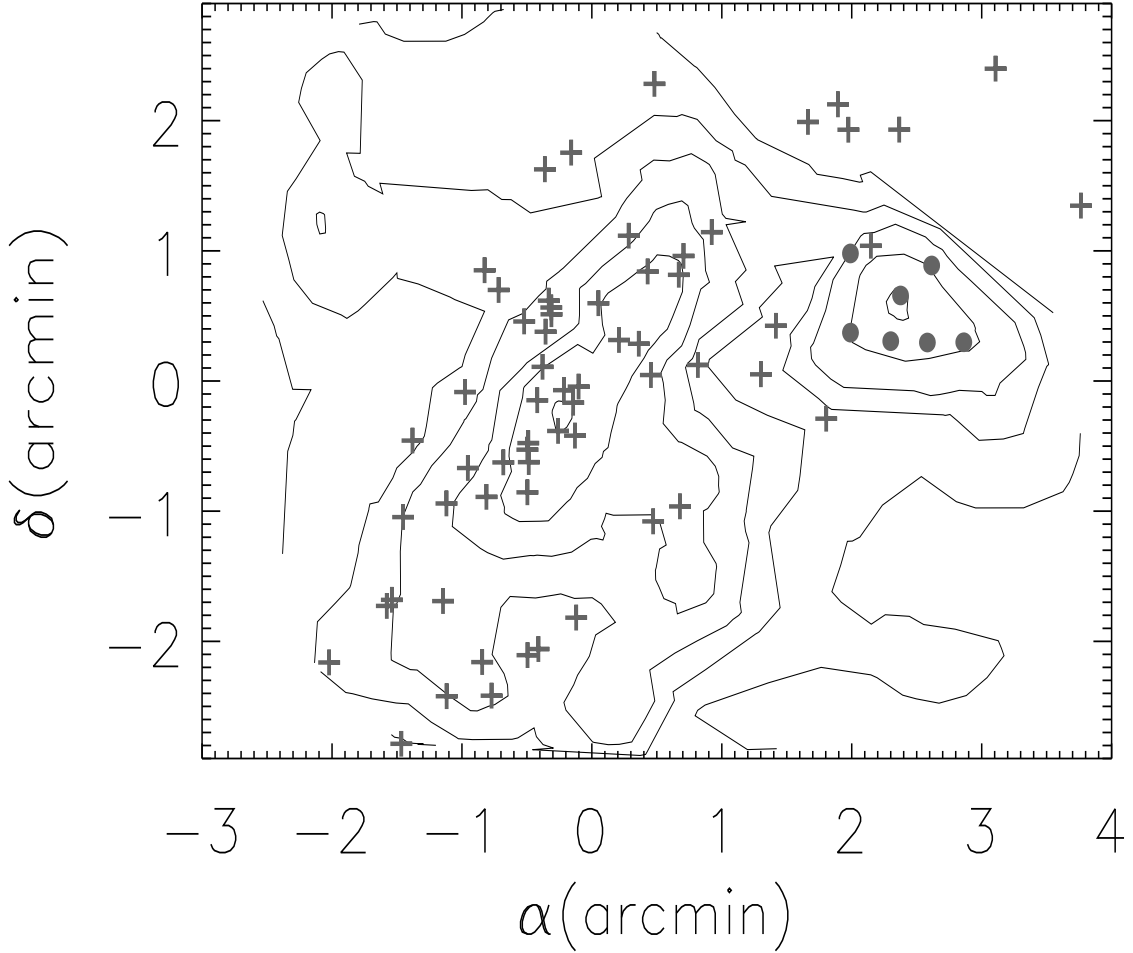}
\caption{2D projected density of (likely) cluster members, selected  
within $\pm 0.5$ mag of the $B-R$ vs. $R$ CM sequence.
The 2D-density is estimated using the Adaptive Kernel Method. 
The three highest-density contours correspond to 
9, 12 and 24$\sigma$ levels. 
North is up, East is to the left. Crosses indicate galaxies
with measured velocities that belong to the main cluster,
dots indicate galaxies with measured velocities that
belong to a subcluster, according to the KMM algorithm
partition (see text).}
\label{gal_dens}
\end{figure*}

We analyze the 2-dimensional (2D) projected distribution of (likely)
cluster members by considering only those galaxies within $\pm 0.5$
mag of the $B-R$ vs. $R$ CM sequence (specifically, those having
magnitudes and colours within the CM band defined by $2.83 \le (B-R) +
0.045 \, R \le 3.83$). The 2D galaxy density distribution, computed with
the Adaptive Kernel Method (e.g. Pisani \cite{pis93}), is shown in
Fig.  \ref{gal_dens}. Two main structures are evident: the elongated
main cluster body, and a roughly circular structure, 0.7 Mpc to the
west, at the cluster distance. Since this additional structure is
populated by galaxies in the CM sequence, it is likely to be a
substructure of the cluster rather than a group in the cluster
foreground or background. The analysis of the spectroscopic sample
confirms that the substructure is roughly at the cluster redshift (see
below).

\begin{figure}
\centering
\includegraphics[width=9cm]{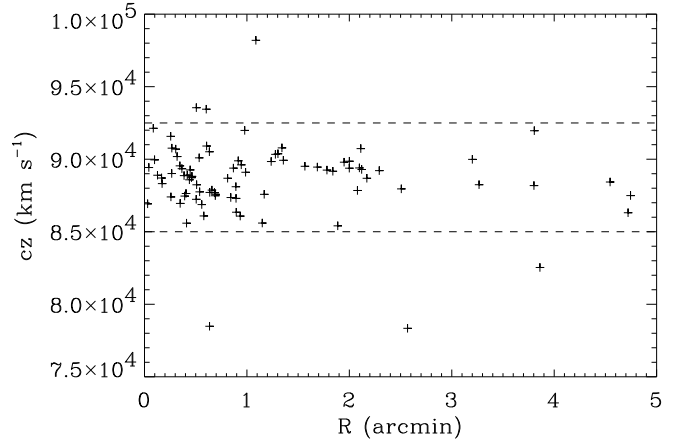}
\caption{The velocity vs. clustercentric distance diagram. The distance
is computed after circularization of the coordinates, using a best-fit
ellipse with an axial ratio of 2, and a position angle of 55$^{\circ}$.
The density peak in the 2D distribution of member galaxies is
taken as the centre of the cluster. The dotted lines indicate the
region of cluster membership in velocity space.}
\label{caustic}
\end{figure}

We use the full spectroscopic sample of 120 galaxies to determine the
cluster membership. The cluster membership is best determined by the
analysis of the caustic diagram in the space of velocities
vs. clustercentric distances (see, e.g., Kent \& Gunn
\cite{ken82}). Since the cluster \mbox{1E0657-56} is significantly
elongated, we first circularize the coordinates. We fit an ellipticity
and an axial ratio to the projected distribution of cluster members
shown in Fig. \ref{gal_dens} (we select only galaxies within the main
cluster body, i.e. excluding the secondary peak to the west). We find
an axial ratio of 2, and a position angle of 55$^{\circ}$. Then we
take the density peak in the 2D distribution of member galaxies as the
centre of the cluster. The caustic diagram is shown in
Fig.~\ref{caustic}.  From the caustic diagram it is straightforward to
choose the range 85000-92500 km~s$^{-1}$ for cluster membership in the
velocity space.  78 galaxies of our spectroscopic sample have measured
velocities within this range. Data for these 78 cluster members are
given in Table~\ref{data}.

The velocity distribution for the 78 cluster members is not
significantly different from a Gaussian (the hypothesis of
Gaussianity is rejected with 1\% probability, according to a
Kolmogorov-Smirnov test, or 11\%, according to a $\chi^2$-test).
Using
the biweight estimator (Beers et al. \cite{bee90}), we compute the
cluster mean heliocentric velocity, $\overline{v}=88777 \pm 63$ km
s$^{-1}$, and its velocity dispersion (in the cluster rest-frame, see
Harrison \& Noonan \cite{har79}) $\sigma_v=1201^{+100}_{-92}$ km
s$^{-1}$ (errors are at the 68~\% confidence level, i.e. 1 $\sigma$
for a normal distribution). The value of the velocity dispersion is in
remarkably good agreement with the preliminary estimate of T98 based
on 13 galaxies, and it is also consistent (within errors) with the
slightly lower estimate by Girardi \& Mezzetti (\cite{gir01}).

\begin{figure}
\centering
\includegraphics[width=9cm]{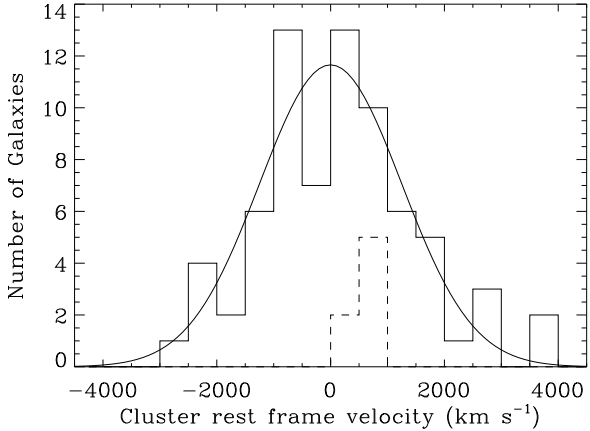}
\caption{The velocity histograms of the 71 cluster members (solid
histogram), and of the 7 subcluster members (dashed histogram). The
Gaussian best-fit to the velocity distribution of the 71 cluster
members is shown.}
\label{histo}
\end{figure}

The gaussianity of the velocity distribution is not sufficient to
exclude the presence of subclustering (see, e.g., Girardi \& Biviano
\cite{gb02}). Detection of substructures is much more efficient when
the full available phase-space information is used.  To this aim, we
apply the KMM algorithm (Ashman et al. \cite{ash94}) to the
distribution of cluster members in 3D-space of positions and
velocities. We search for the solution that separates the cluster
members into two systems. The KMM algorithm makes use of the
Maximum-Likelyhood Ratio test to estimate how likely the two-system
solution is to be a significant improvement over the single-system
solution. In our case, the two-system solution is significantly better
than the single-system solution, at the 99.9\% confidence level
(c.l.).  KMM assigns 7 galaxies to the secondary system, each with a
$\geq 99.9$\% c.l.  From these galaxies, we compute the subcluster
mean velocity $\overline{v}=89479 \pm 41$ km s$^{-1}$, and its
velocity dispersion, $\sigma_v=212^{+67}_{-52}$ km s$^{-1}$. The mean
velocity and velocity dispersion of the main system are almost
unchanged when the 7 galaxies belonging to the subcluster are removed
from the sample of cluster members ($\overline{v}=88681 \pm 69$ km
s$^{-1}$ and $\sigma_v=1249^{+109}_{-100}$ km s$^{-1}$). The
subcluster is significantly offset in velocity from the main system,
$\Delta \overline{v} = 616 \pm 80$ km~s$^{-1}$, in the cluster rest
frame.  In Fig. \ref{histo} we show the two velocity histograms of the
71 cluster members, and of the 7 subcluster members, as well as the
Gaussian distribution that best fits the velocity histogram of cluster
members.  In Fig.~\ref{gal_dens} we overplot the galaxies with
measured velocities, belonging to the main cluster (crosses) and to
the subcluster (squares), onto the 2D map of projected density counts.

\subsection{Virial mass estimates}
\label{virial}
From the substructure analysis, we identify 7 galaxies belonging to a
system separate from the main cluster. The virial mass of the main
cluster computed on the 71 remaining cluster members, is $M_{\rm{vir}}
= 1.33 \times 10^{15} \, \mathrm{M}_{\odot}$, with a 10~\% uncertainty
(estimated with the jackknife technique -- see, e.g., Beers et al.
\cite{bee84}).  The virial mass is estimated within an aperture of
$\sim 1.5$~Mpc, which is only $\sim 60$\% of the cluster virial radius
computed as $r_{200}=\sqrt{3} \sigma_v / (10 H_{\rm{z}})$ (Carlberg et
al. \cite{car97}).  We therefore need to correct this mass for the
surface-pressure term (The \& White \cite{the86}). We apply this
correction following the procedure of Girardi et
al. (\cite{gir98}). The corrected virial mass is $M_{\rm{vir,c}}(<0.6
r_{200}) = 0.82 \times 10^{15} \, \mathrm{M}_{\odot}$.  We extrapolate
this mass to $r_{200}$, $M_{\rm{vir,c}}(<r_{200}) = 1.24 \times
10^{15} \, \mathrm{M}_{\odot}$, assuming a NFW (Navarro et
al. \cite{nav97}) mass profile with a scale $0.2 \, r_{200}$. This
value for the cluster mass is in agreement (within errors) with that
derived by Girardi \& Mezzetti (\cite{gir01}).

Our estimate of the virial mass for the subcluster is $M_{\rm{vir}} = 0.13
\times 10^{14} \, \mathrm{M}_{\odot}$, with an uncertainty of 27~\%. Applying the
surface-term correction, and extrapolating to the subcluster
$r_{200}$ as we did for the main cluster, leads the virial
estimate to $M_{\rm{vir,c}}(<r_{200}) = 0.12 \times 10^{14} \, \mathrm{M}_{\odot}$.

How reliable are these mass determinations? We argue below (see
Sect.~\ref{discuss}) that we are observing \mbox{1E0657-56} soon after
the collision with a cluster of intermediate mass (or group). The main
cluster dynamics does not seem to have been substantially affected by
this collision (see Sect.~\ref{discuss}) so that the (corrected and
extrapolated) virial mass estimate of the main system is probably
reliable. On the other hand, the subcluster is likely to have been
partly disrupted by the collision, and it is difficult to estimate its
mass based on equilibrium models. In these conditions, neither the
application of the virial theorem, nor the surface-pressure term
correction and extrapolation to $r_{200}$ are warranted. Taking into
account these systematic uncertainties, we determine a fiducial mass
range for the subcluster of $0.07$--$0.34 \times 10^{14} \,
\mathrm{M}_{\odot}$.

\begin{figure}
\centering
\includegraphics[width=9cm]{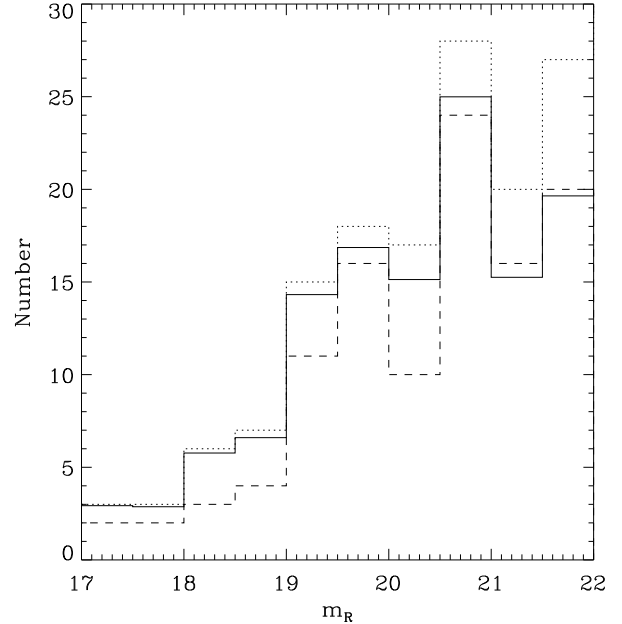}
\caption{Histograms of the $R$-band apparent magnitudes of galaxies
within a 5.4 arcmin$^2$ cluster region covered by our photometric
observations, excluding the subcluster region. The dotted-line
represents counts of all galaxies in the selected region,
the solid-line the cluster counts, after field count
subtraction. The dashed line represents all galaxies within
a band $\pm 0.5$ mag centred on the cluster CM sequence.}
\label{histolum}
\end{figure}
\subsection{Mass-to-light ratios}
\label{mlr}
To compute the luminosity of the cluster, we consider
galaxies within a 5.4 arcmin$^2$ region which excludes the
subcluster. We then follow two alternative approaches.  In the first
approach we consider all galaxies with magnitude $17 \leq R \leq 22$,
and subtract the field counts, taken from the literature (Roche et
al. \cite{roc96}; Shanks et al. \cite{sha84}; Weir et
al. \cite{wei95}). We choose the magnitude range so as to avoid major
background and foreground contamination.  In the second approach we
consider all galaxies within $\pm 0.5$ mag around the cluster CM
sequence (see Sect.~\ref{galdis}), within the same magnitude range as
above. In this case, the field contribution is statistically
eliminated by the selection of galaxies in the CM diagram. As can be
seen from Fig. \ref{histolum}, the two approaches lead to very similar
magnitude distributions of (likely) cluster members.

\begin{figure*}
\centering
\includegraphics[width=\textwidth]{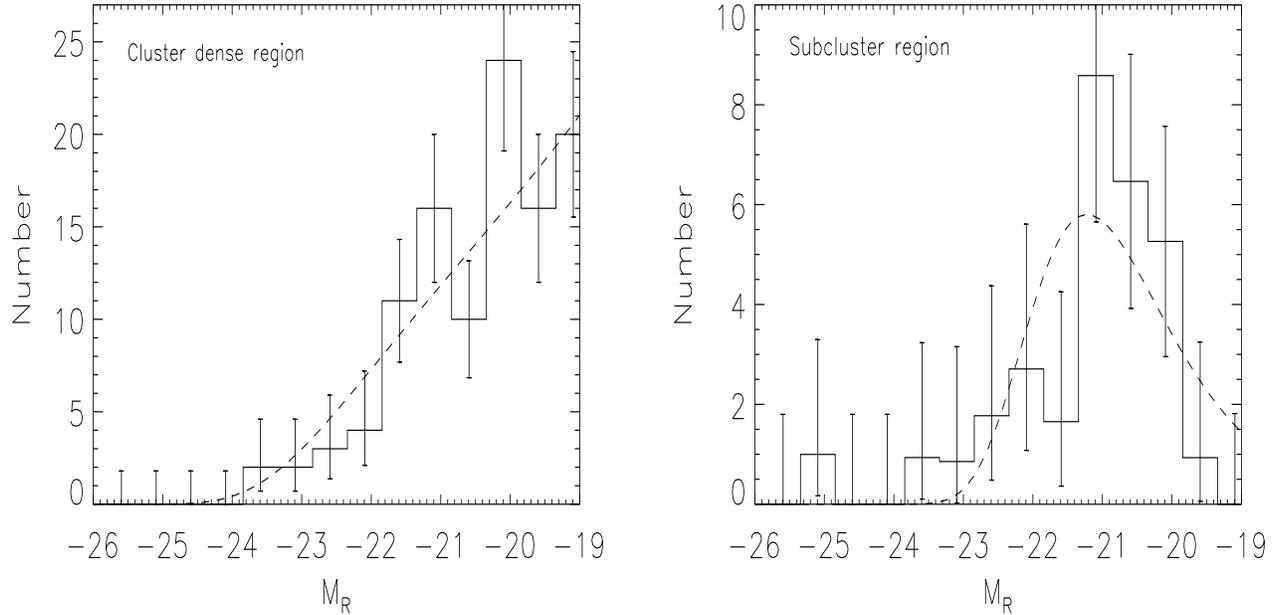}
\caption{$R$-band luminosity functions for
the cluster and subcluster regions. 1-$\sigma$ error bars
are shown. Dashed lines show best-fits with Schechter functions.}
\label{lumfun}
\end{figure*}

We compute absolute magnitudes using the cluster distance-modulus and
the evolutionary- and K-correction in the $R$ band for an early-type
galaxy at the redshift of \mbox{1E0657-56} (Poggianti \cite{pog97}).
We then fit a Schechter (\cite{sch76}) function to the absolute
magnitude distributions obtained as described in the previous
paragraph. The best fit parameters are $M_{\rm{R}}^{\star}=-23.01 \pm 0.13$,
$\alpha=-1.23 \pm 0.10$ for the field-subtracted counts. We obtain the
same result, within errors, for the CM band-selected galaxies (see
Fig.~\ref{lumfun}). Integration of the best-fit Schechter luminosity
function over the magnitude range $-24 \leq R \leq -14$ yields a total
luminosity of $L_{\rm{R}} = 1.00 \times 10^{12} \, \mathrm{L}_{\odot}$, $\pm
14$ \% depending on the adopted method.  The (virial) mass-to-light
ratio within the same region is $M/L_{\rm{R}} = 199 \pm 29 \,
\mathrm{M}_{\odot}/\mathrm{L}_{\odot}$. The mass-to-light ratio of
\mbox{1E0657-56} is consistent with the value
$M/L_{\rm{R}} = 217 \, \mathrm{M}_{\odot}/\mathrm{L}_{\odot}$, the mean
mass-to-light ratio obtained for CNOC clusters (Carlberg et
al. \cite{car97b}), once their luminosities are similarly corrected
for passive evolution, and transformed from Gunn $r$ to Bessel $R$.

We estimate in a similar way the total luminosity of the subcluster.
We take the observed galaxy counts in a 1.3 arcmin$^2$ region centered
on the subcluster, and subtract the field counts plus the estimated
contribution from the cluster at the subcluster distance. The
resulting absolute magnitude distribution has a Gaussian-like shape
(see Fig.~\ref{lumfun}) and is markedly different from the
magnitude distribution of cluster galaxies.  The best-fit Schechter
parameters of the subcluster magnitude distribution are rather poorly
constrained. However, given the lack of faint galaxies, there is no
need for extrapolation of the luminosity function to faint
magnitudes. We can therefore estimate the total subcluster luminosity
simply from the observed distribution. We thus obtain $L_{\rm{R}} = 0.2
\times 10^{12} \, \mathrm{L}_{\odot}$.  The resulting mass-to-light
ratio is $M/L_{\rm{R}} = 35$--$170 \, \mathrm{M}_{\odot}/\mathrm{L}_{\odot}$
(see Sect.~\ref{virial}). The upper limit of the subcluster
mass-to-light ratio is consistent with the mass-to-light ratio found
for the main cluster.  The lower limit is consistent with the first
quartile of the distribution of mass-to-light ratios found for loose
groups (Ramella et al. \cite{ram89}), after transformation from the
blue to the red band.  Low mass-to-light ratios are also found at the
centre of rich clusters, where luminosity segregation has occurred
(see, e.g., Koranyi et al. \cite{kor98}).

\subsection{Two-body dynamical model}
\label{bimo}
\begin{figure*}
\centering
\includegraphics[width=8cm]{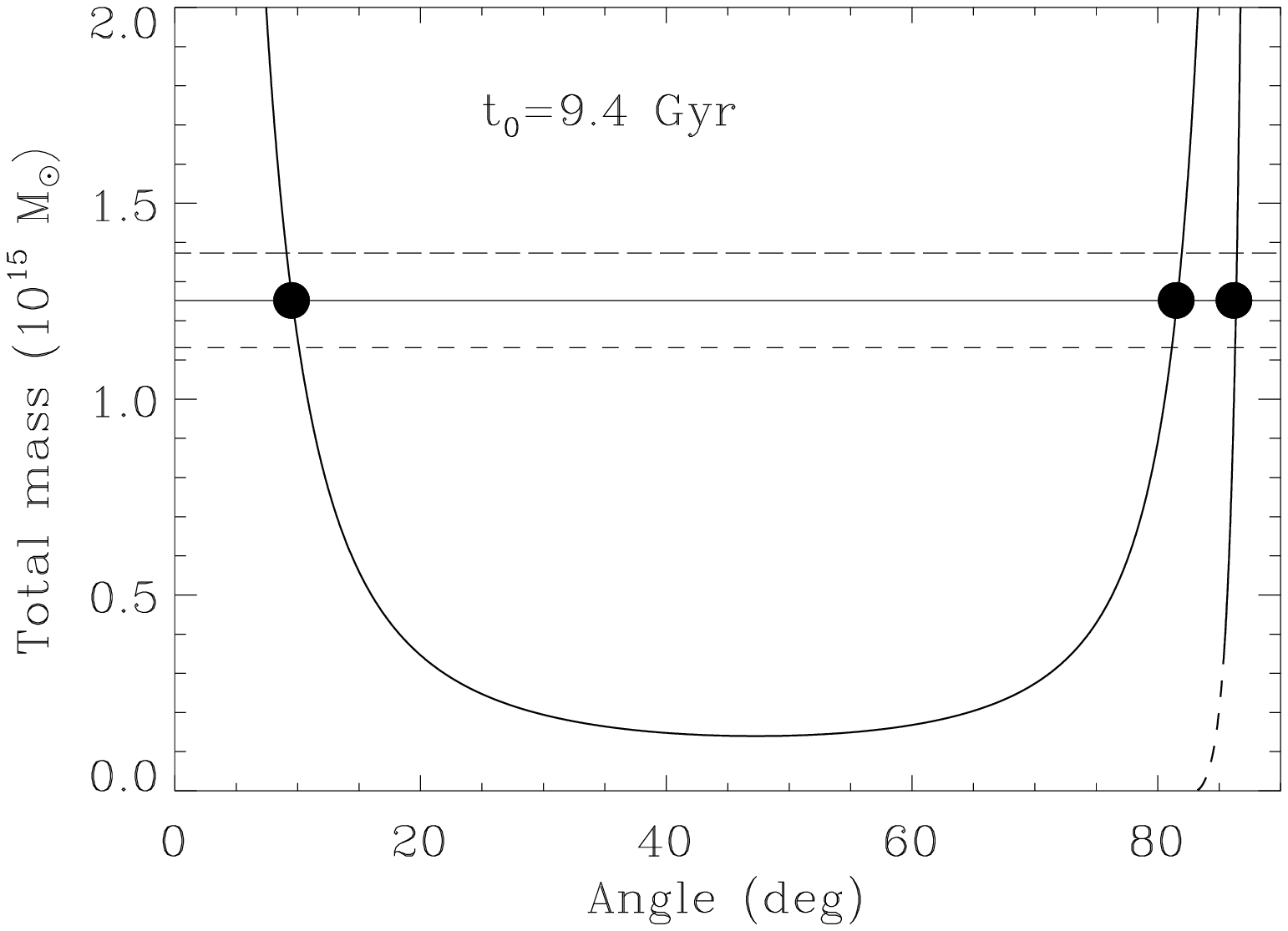}
\includegraphics[width=8cm]{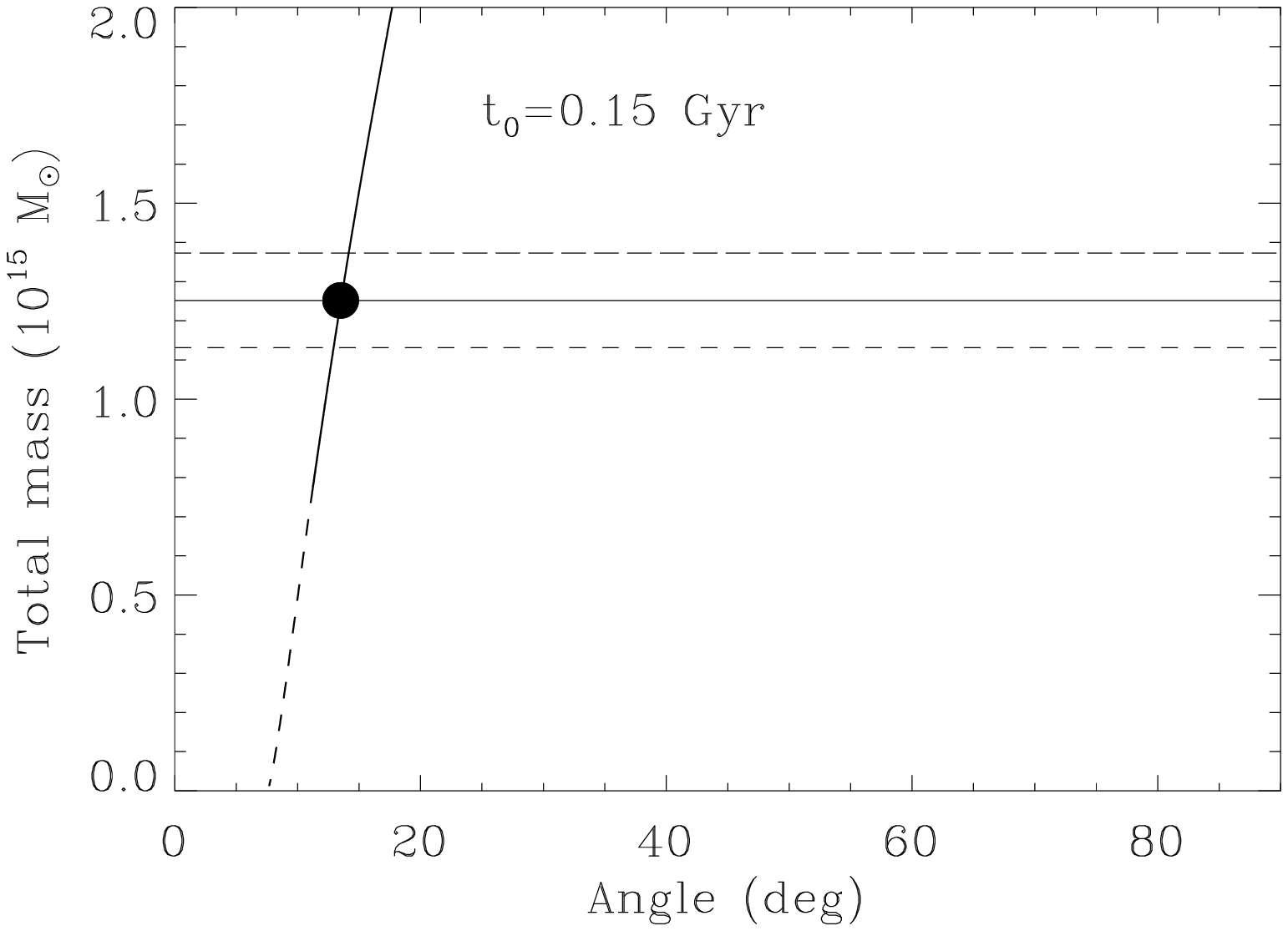}
\caption{The sum of the virial masses of the main cluster and its
subcluster as a function of the angle between the plane of the sky and
the line connecting the two systems, in the two-body dynamical model.
The two systems have a projected distance of 0.7 Mpc and a radial
velocity difference of 616 km~s$^{-1}$ in the cluster rest frame.  The
horizontal lines represent the sum of the virial masses of the two
systems, and its confidence band.  {\bf Left panel:} The solid lines
show the bound-incoming and the bound-outgoing solutions. The dashed
line shows the unbound solution. These solutions are derived assuming
that the two systems were at zero separation 9.4 Gyr ago, the age of
the Universe at $z \simeq 0.3$ in our adopted cosmology, and since
then, they have never been in contact again. The three possible bound
configurations for the \mbox{1E0657-56} cluster and its subcluster are
indicated with filled circles.  {\bf Right panel:} The solid line
shows the bound-outgoing solution.  The dashed line shows the unbound
solution.  These solutions are derived assuming that the two systems
were at zero separation 0.15 Gyr ago. In this case, the only
acceptable solution for the system mass corresponds to the
bound-outgoing case (indicated with a filled circle).}
\label{bimodal}
\end{figure*}

Using the virial masses of the two galaxy systems, their projected
distance, and their relative velocity along the line-of-sight (see
Sect.~\ref{galdis} and \ref{virial}), we carry out the dynamical
analysis of the cluster+subcluster system, with the two-body model
(Gregory \& Thompson \cite{gre84}; Beers et al. \cite{bee91}).

In Fig. \ref{bimodal} we show the modeled total mass of the cluster
plus subcluster vs. the angle between the plane of the sky and the
line connecting the two systems. In the left panel, we show the
solution for the case in which the cluster and subcluster are
expanding or approaching for the first time. In this case we assume
that the two systems were at zero separation 9.4 Gyr ago, which is the
age of the Universe at the cluster redshift in our adopted cosmology.
From this model, we conclude that the subcluster is bound to the
system. The two systems could be currently approaching, or they could
still have to reach maximum expansion, depending on the unknown
geometry of the collision.

Based on {\em Chandra} observations, M01 have recently suggested that
a collision between the subcluster and the main cluster has already
occurred. From the density jump at the X-ray shock front, M01 estimate
that the subcluster is currently moving away from the cluster with a
velocity of 3000-4000 km~s$^{-1}$. From the observed line-of-sight
component of the relative velocity between the cluster and the
subcluster (see Sect.~\ref{galdis}) we infer a projection angle of
5--15 degrees between the line connecting the two systems and the
plane of the sky (i.e. the subcluster is moving nearly in the plane of
the sky). M01 reach the same conclusion based on the sharpness of the
X-ray brightness edge that they identify as a bow shock.

Assuming that the cluster and the subcluster have already crossed each
other, we determine different solutions for the two-body model for
different times of the collision event. Only by setting the collision
epoch $\simeq 0.15$ Gyr ago, does a solution exist in which the
subcluster velocity is as predicted by the X-ray analysis of M01. This
solution (shown in Fig. \ref{bimodal}, right panel) implies that the
subcluster is currently moving away from the main cluster, and it will
eventually collide again in the future, after reaching maximum
expansion at about twice the current distance from the cluster
centre. For collision epochs $\leq 0.1$ Gyr or $\geq 0.2$ Gyr, the
current subcluster velocity would be significantly larger, or,
respectively, smaller, than predicted by the X-ray analysis of M01.

\subsection{Different cluster populations}
\label{morph}
It has been suggested that cluster-cluster collisions may trigger
star formation in cluster galaxies (Bekki \cite{bek99}; Moss \&
Whittle \cite{mos00}; Girardi \& Biviano \cite{gb02} and references
therein). Caldwell \& Rose (\cite{cal97}) noticed that post-starburst
galaxies are frequently found in clusters with evidence of past
collision events. Bardelli et al. (\cite{bar98}) found that the bluest
galaxies in the ABCG~3558/3562 supercluster are located in the region
between the two colliding clusters. Here we consider the relative
fractions and distributions of galaxies of different colours, spectral
and/or morphological types in \mbox{1E0657-56}.

Since the fraction of blue galaxies in clusters may depend on redshift
(the so-called ``Butcher-Oemler'' effect, Butcher \& Oemler
\cite{but78}, \cite{but84}) it is important to compare the blue galaxy
fraction in \mbox{1E0657-56} with the mean blue galaxy fraction of
other clusters at the same redshift. We estimate the blue galaxy
fraction in \mbox{1E0657-56} in two ways: (a) using our photometric
sample, according to the definition of Margoniner et
al. (\cite{mar01}, MCGD hereafter), and (b) using our spectroscopic
sample, according to the definition of Ellingson et al. (\cite{ell01},
ELYC hereafter).  In Table~\ref{popfrac} we list the resulting blue
galaxy fractions in \mbox{1E0657-56}, and, for comparison, the mean
blue galaxy fractions found by MCGD and ELYC for clusters at $z \simeq
0.3$. The \mbox{1E0657-56} blue galaxy fraction as derived on
the photometric sample is marginally larger than the mean found by
MCGD for $z \simeq 0.3$ clusters, but the difference is not
significant.  From the values listed in Table~\ref{popfrac} we
conclude that the blue galaxy fraction of \mbox{1E0657-56} is
consistent with those of $z \simeq 0.3$ clusters.

\begin{table}
\centering
\caption[]{Population fractions}
\label{popfrac}
\begin{tabular}{cll}
\hline
\hline
Cluster population & \mbox{1E0657-56} & Literature \\
\hline
Blue -- photom. sample & $0.53 \pm 0.11$ & 0.36 (MCGD) \\
Blue -- spectro. sample & $0.14 \pm 0.05$ & 0.15 (ELYC)\\
Elliptical-like spectra & $0.75 \pm 0.14$ & 0.54 (ELYC)\\
                        &                 & 0.47 (D99)\\
Balmer-absorption spectra  & $0.22 \pm 0.06$ & 0.30 (ELYC)\\
                           &                 & 0.20 (D99) \\
Emission-line spectra & $0.03 \pm 0.02$ & 0.16 (ELYC) \\
                      &                 & 0.32 (D99) \\
\hline                                                           
\end{tabular}
\end{table}

We then compare the fractions of different spectral-type populations
in \mbox{1E0657-56} with those determined by ELYC on the CNOC
clusters, and by Dressler et al. (\cite{dre99}, hereafter D99) on the
MORPHS clusters.  ELYC consider three spectral-types: elliptical-like
spectra, spectra with strong Balmer-absorption, and spectra with
emission lines (emission-line galaxies, ELG, hereafter). A
finer classification scheme was devised by D99, depending on the
EW(H$\delta$) and the EW of emission lines.  For lack of statistics
(we only have spectral types for 67 cluster members in total), we
prefer to join together some of their classes.  In particular, we
combine the $k+a$ and $a+k$ classes of D99 into a single class,
corresponding to ELYC's class of spectra with strong Balmer
absorption. Similarly, we combine D99's $e(a)$, $e(b)$, and $e(c)$
classes into a single class, which corresponds to ELYC's class of
emission-line galaxies.  D99's $k$ class corresponds to ELYC's
elliptical-like spectral class.

In Table~\ref{popfrac} we list the fractions of the different
spectral-type galaxies in \mbox{1E0657-56}, and, for comparison, the
mean fractions found by ELYC in the CNOC clusters, and by D99 in the
MORPHS clusters. The fraction of galaxies with strong or moderate
Balmer-absorption in \mbox{1E0657-56} is similar to those found by ELYC
and D99. On the other hand, in \mbox{1E0657-56} there is a lower
fraction of galaxies with emission-lines, and a higher fraction of
galaxies with elliptical-like spectra, than the average fractions
found by ELYC and D99.

\begin{figure}
\centering
\includegraphics[width=9cm]{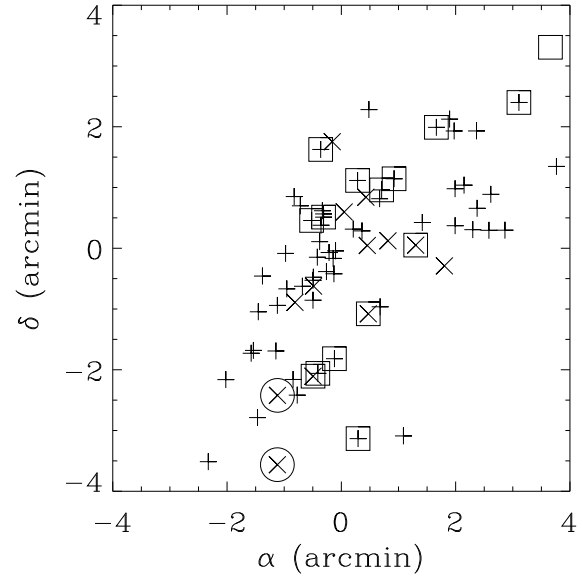}
\caption{Spatial distributions of early-type galaxies (crosses),
late-type galaxies (X's), galaxies with EW(H$\delta$) $\geq 3$~\AA
(squares) and galaxies with emission lines (circles). Only galaxies
with velocities in the cluster velocity range are shown.}
\label{specspat}
\end{figure}

\begin{figure}
\centering
\includegraphics[width=9cm]{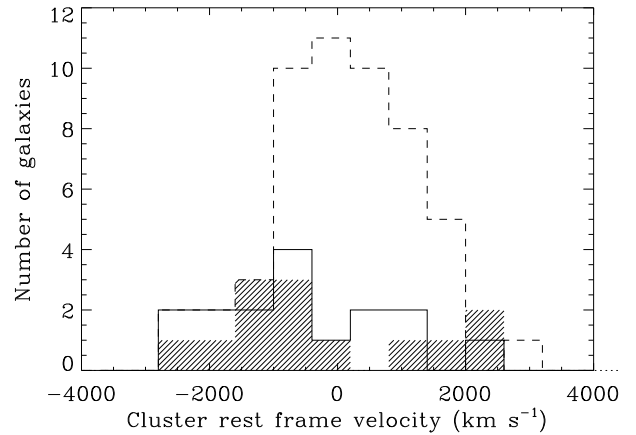}
\caption{Velocity distributions of early-type galaxies (dashed line), 
late-type galaxies (shaded histogram), galaxies with
EW(H$\delta$) $\geq 3$~\AA~ (solid line), in the reference frame of
the cluster.  Only cluster members are considered; galaxies belonging
to the subcluster have been removed from the sample.}
\label{velpop}
\end{figure}

We now consider the distributions of the different cluster
populations.  In Fig.~\ref{specspat} we show the spatial distributions
of different populations of cluster members: galaxies classified
morphologically as early-type and late-type, galaxies with
EW(H$\delta$) $\geq 3$~\AA, and galaxies with emission lines. The
early-type galaxies seem to be more centrally concentrated than the
other galaxy populations. A Rank-Sum test (e.g. Hoel \cite{hoe71})
confirms this visual impression (95\% c.l.). The only two
emission-line galaxies are located in the cluster outskirts, similarly
to what is usually seen in nearby clusters (Biviano et
al. \cite{biv97}).  We show in Fig.~\ref{velpop} the velocity
distributions of the different cluster populations (after getting rid
of galaxies in the subcluster), and we list their mean velocities and
velocity dispersions in Table~\ref{popvel}. The velocity dispersion of
the early-type galaxies seems to be lower than that of the late-type
galaxies, but the difference is not significant (according to the
F-test, see, e.g., Press et al. \cite{press}). The only significant
difference (at the 95~\% c.l. according to a Rank-Sum test) occurs
between the mean velocities of the early-type galaxies and of the
strong Balmer-absorption galaxies.

\begin{table}
\centering
\caption[]{Velocity distributions of different populations}
\label{popvel}
\begin{tabular}{clll}
\hline
\hline
Cluster population  & $N_{\rm{gal}}$ & $\overline{v}$ & $\sigma_v$ \\
& & km s$^{-1}$ & km s$^{-1}$ \\
\hline
early-type & 52 & $88765 \pm 75$ & $1132_{-106}^{+117}$ \\
late-type & 13 & $88324 \pm 227$ & $1682_{-306}^{+371}$ \\
Balmer-absorption & 16 & $88097 \pm 174$ & $1434_{-236}^{+281}$ \\
\hline                                                           
\end{tabular}
\end{table}

Finally, we note that all 7 spectroscopically-confirmed members of
the subcluster can be assumed to be evolved ellipticals, or S0--Sa's,
since we classify them morphologically as early-type, and
spectroscopically as $k$-type.

\begin{figure*}
\centering
\includegraphics[width=8cm]{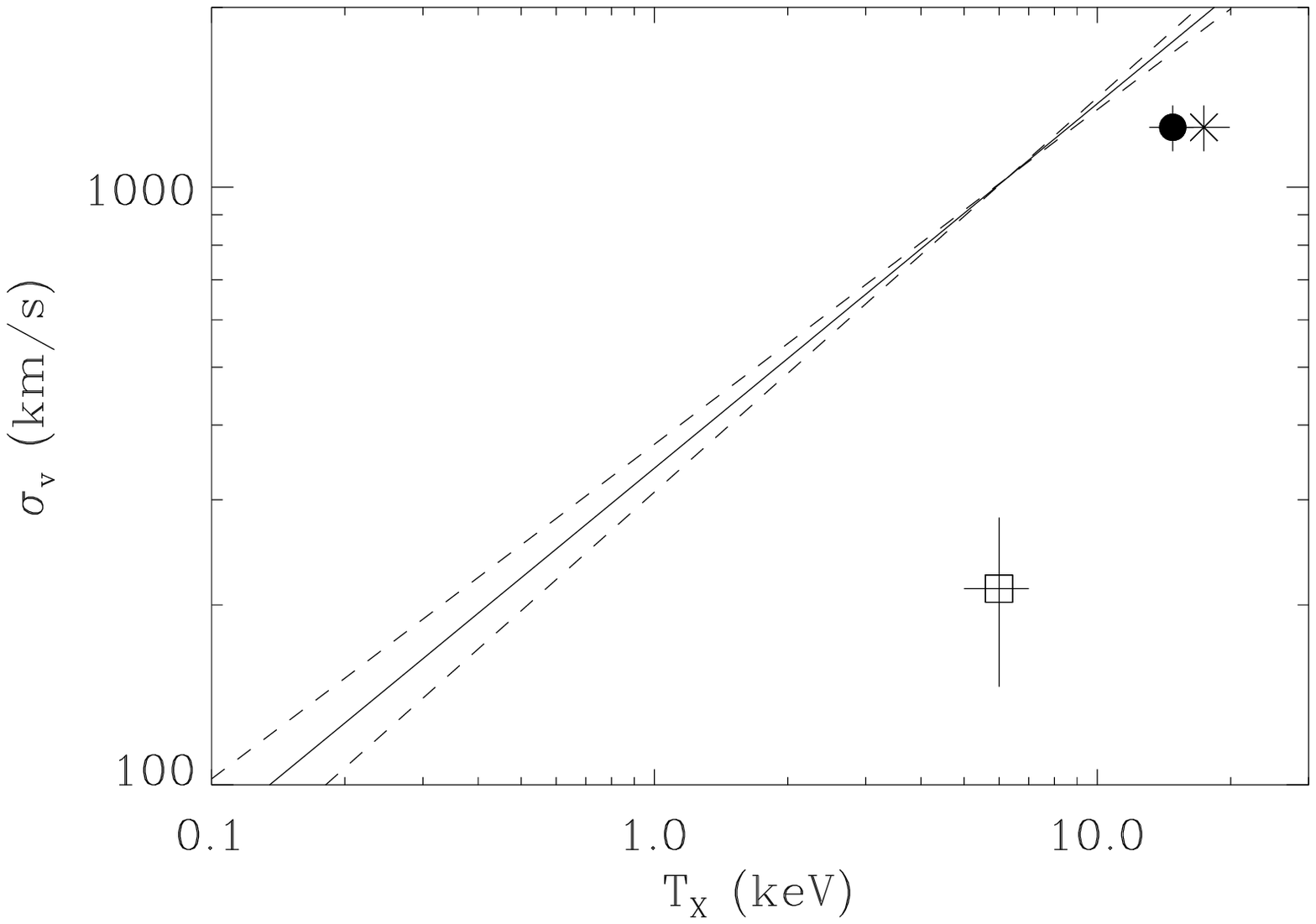}
\includegraphics[width=8cm]{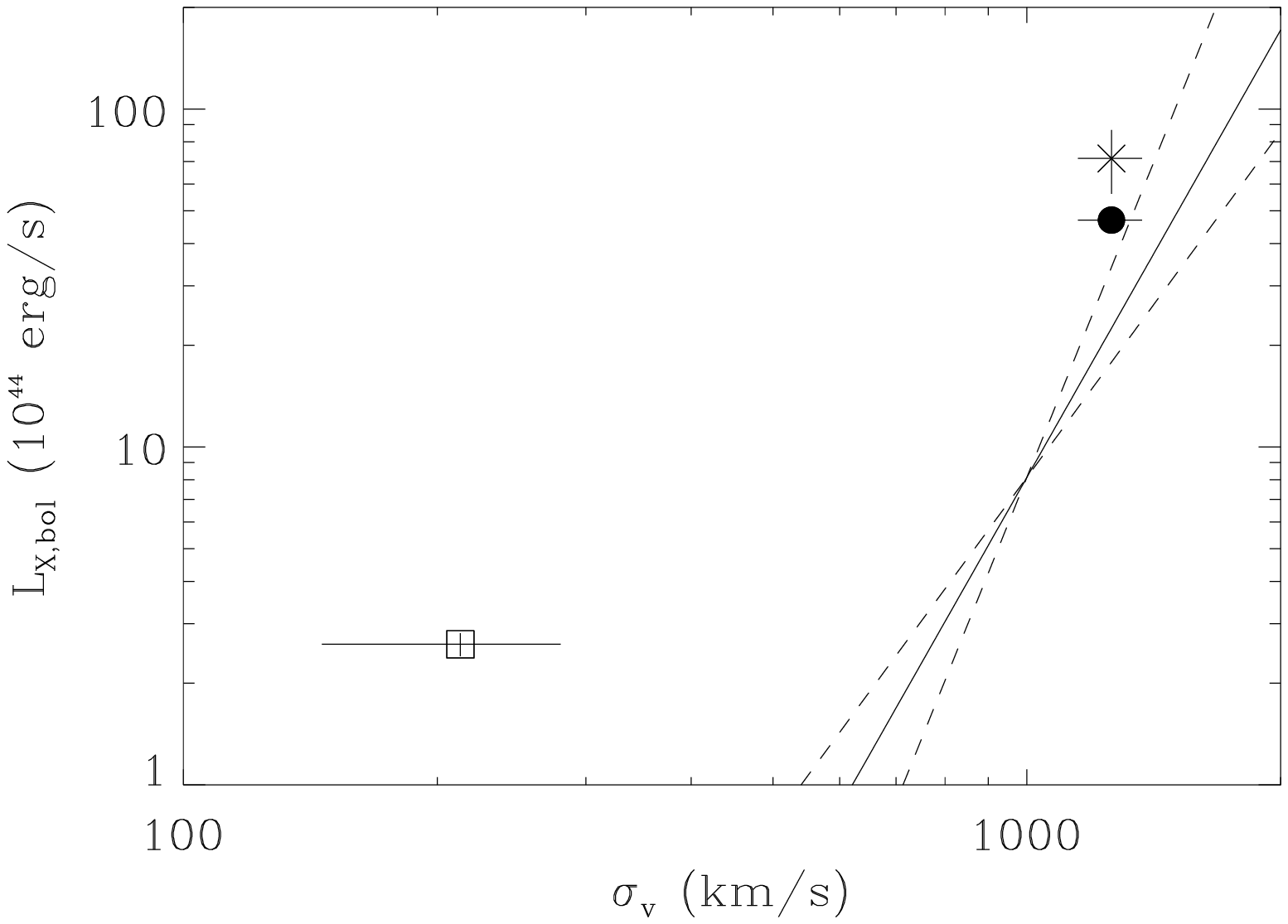}
\caption{{\bf Left panel:} The velocity dispersion vs. X-ray
temperature relation of Girardi et al. (\cite{gir96}; solid line),
within its $\pm 1 \sigma$ relations (dashed lines). The locations of
the subcluster (square) and the cluster (dot: values from M01; X:
values from T98) are shown.  {\bf Right panel:} The X-ray luminosity
vs. velocity dispersion relations of Girardi \& Mezzetti
(\cite{gir01}; solid line) within its $\pm 1 \sigma$ relations (dashed
lines).  The locations of the subcluster (square) and the cluster
(dot: values from M01; X: values from T98) are shown.}
\label{svx}
\end{figure*}

\section{Discussion}
\label{discuss}
\subsection{A major collision event?}
Using our new spectroscopic sample of \mbox{1E0657-56} members in a
$\sim 1.8$ Mpc$^2$ region, we detect a subcluster of low velocity
dispersion, $\sigma_v \simeq 200$ km s$^{-1}$, 
$\simeq 600$ km~s$^{-1}$ and 0.7 Mpc away from the main cluster (see
Sect.~\ref{galdis}). This subcluster was also recently detected in
the X-ray by M01 using {\em Chandra.} M01 suggest that a bow shock is
currently propagating outward from the cluster, nearly at the location
of our optically detected subcluster.  They determine a shock velocity
of 3000-4000 km~s$^{-1}$. Assuming that the subcluster galaxies are
also moving away with this velocity, we determine the projection angle
of the cluster-subcluster system from the observed line-of-sight
velocity of the subcluster relative to the main cluster. We find that
the line connecting the two systems lies very close to the plane of
the sky. With such an orientation angle, the two-body dynamical model
for the cluster and its subcluster has two solutions (see
Sect.~\ref{bimo}). In both solutions, the two systems are
gravitationally bound. If we further assume that the subcluster is
moving away from the main cluster, we can constrain the epoch of
collision to $\simeq 0.15 \pm 0.05$ Gyr ago. The subcluster will reach
maximum expansion at $\sim 1.2$ Mpc away from the main cluster, and
then recollapse again.

The (virial) mass ratio of the subcluster and the main cluster is very
low, ranging from 1:200 to 1:30 (see Sect.~\ref{virial}).  The allowed
range is large because of the very uncertain mass of the
subcluster. The low velocity dispersion of the subcluster ($\sim 200$
km~s$^{-1}$) is typical of loose groups (e.g. Ramella et
al. \cite{ram89}), but also of the cores of rich clusters that have
developed luminosity segregation (Biviano et al. \cite{biv92}). The
nearly Gaussian shape of the subcluster magnitude distribution
resembles that of cluster cores, where dynamical friction and merging
of bright galaxies, together with tidal disruption of faint galaxies,
tend to enrich the bright end of the magnitude distribution at the
expense of the faint end. Cluster cores are dominated by early-type
galaxies, the same galaxy population which characterizes the
subcluster. These considerations suggest that the subcluster could be
the remnant core of a moderately massive cluster disrupted by its
collision of \mbox{1E0657-56}. In this case, the pre-merger mass ratio
could have been substantially larger.

\begin{figure*}
\centering
\includegraphics[width=\textwidth]{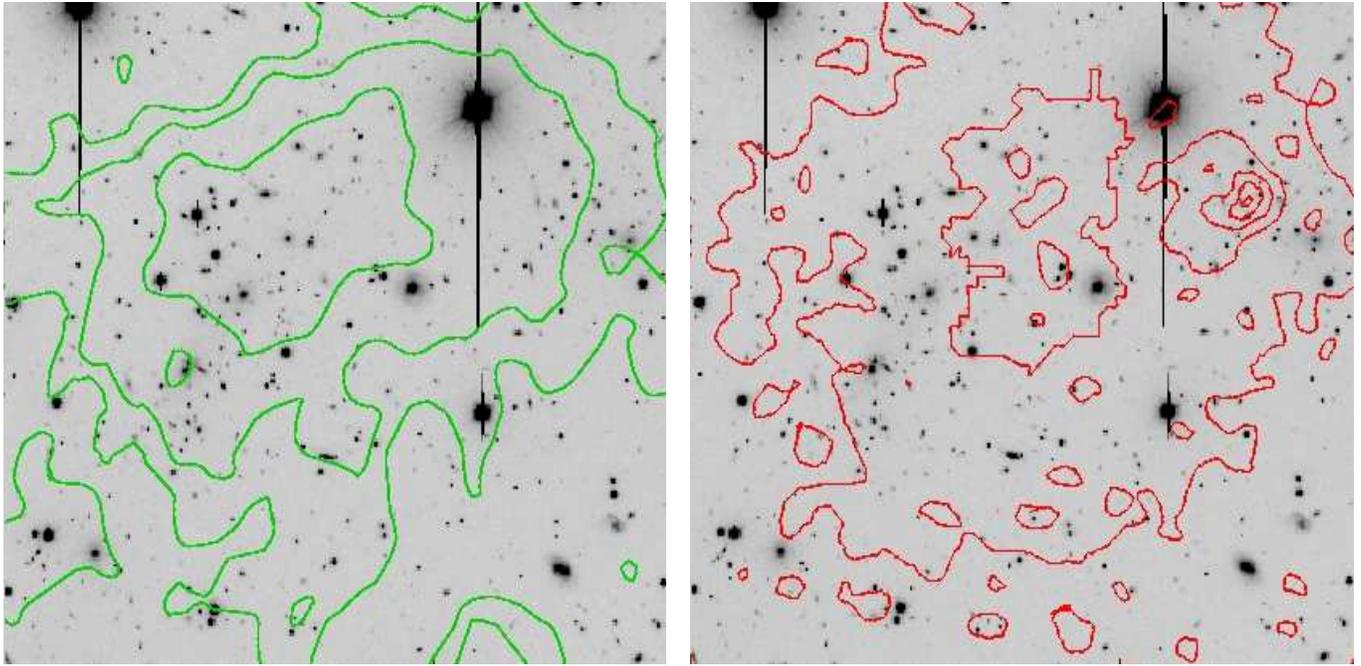}
\caption{{\bf Left panel:} contours of the 1.3 GHz radio image (after
subtraction of discrete sources and smoothing with a Gaussian filter
of 16$''$ FWHM), from Liang et al. (\cite{lia00}), superposed onto the
$R$-band FORS/VLT image of 1ES0657-056. Contour levels are 5, 12 and
30 $\sigma$, where $\sigma$ correspond to 20 $\mu$Jy$/$beam. {\bf
Right panel:} contours of ROSAT/HRI image smoothed with a 90$''$
Gaussian filter, superposed on the same $R$-band image. The X-ray
contours correspond to 3, 7, 12, 16 and 20$\sigma$ levels above the
background. The scale of both figures is $4.5' \times 4.5'$. North is
up and East is to the left.}
\label{rad_x}
\end{figure*}

Further evidence for this scenario is given by the comparison of the
optical and X-ray properties of \mbox{1E0657-56} and its
subcluster. The subcluster X-ray temperature and luminosity are too
high for its velocity dispersion, as implied by the empirical
relations for galaxy systems of Girardi et al. (\cite{gir96}), and
Girardi \& Mezzetti (\cite{gir01}) (see Fig.\ref{svx}). What we now
identify as the subcluster in the galaxy distribution could be the
remnant compact core of a more massive system that has lost most of
its galaxies after the collision. Similar characteristics of
compactness are also seen in the X-ray emissivity contours, resembling
cooling flow regions in cluster cores (M01). The X-ray temperature and
luminosity of the subcluster corresponds to a velocity dispersion of
$\sim 700$ km~s$^{-1}$ and a pre-merger subcluster-cluster mass ratio
of $\sim$~1:6.

A large mass ratio merger is also suggested by the comparison of the
X-ray and optical properties of the main cluster. The bulk X-ray
emission of \mbox{1E0657-56} is offset from the main concentration of
cluster galaxies, in the direction of the subcluster. This is clearly
seen in the {\em Chandra} image reproduced in Fig.1b of M01, as well
as in the {\em ROSAT/HRI} image\footnote{This image is retrieved from
the {\em ROSAT} public archive at
http://heasarc.gsfc.nasa.gov/docs/rosat/rhp\_archive.html.} shown in
Fig.~\ref{rad_x} (right panel), both over-plotted on the $R$-band FORS
image of the cluster. The X-ray cluster morphology resembles those
seen in the numerical simulations of Roettiger et al. (\cite{roe93},
\cite{roe96}) soon after the collision has occurred. Remarkably, in
these simulations the dark matter distribution of the main cluster is
much less affected by the collision than the gas distribution
(Roettiger et al. \cite{roe93}).  Being a nearly collisionless
component, galaxies probably trace a distribution similar to that of
the dark matter. The observed offset between the galaxy and gas
distributions, is then an expected consequence of a merger, but only
for large mass ratios between the merging systems (compare Fig.1c to
Fig.2c in Roettiger et al. \cite{roe96}, corresponding to the X-ray
surface brightness images of a 1:4 and a 1:8 merger, respectively,
both 0.5 Gyr after core passage).

According to numerical simulations, cluster-cluster collisions not
only affect the X-ray surface brightness distribution, but also
increase the X-ray temperature, X-ray luminosity (Ricker \& Sarazin
\cite{ric01}), and velocity dispersion (Pinkney et al. \cite{pin96})
of the main cluster.  However, the increase in the observed velocity
dispersion depends on the relative orientation between the
line-of-sight and the merger plane. When this is nearly 90 degrees, as
in the case of \mbox{1E0657-56} (see Sect.~\ref{bimo}), the increase
is only $\sim 20$\% for a 1:3 mass ratio (Pinkney et
al. \cite{pin96}).  On the other hand, in the same collision, the
X-ray luminosity and temperature can be boosted up by factors as high
as $\sim 4$ and $\sim 2$ (Ricker \& Sarazin \cite{ric01}). These
theoretical results are consistent with our observations. Both the
X-ray temperature and X-ray luminosity of \mbox{1E0657-56} appear too
high for the cluster velocity dispersion (see Fig.\ref{svx}). On the
other hand, the line-of-sight component of the cluster velocity
dispersion should not have been affected by the collision, or the
cluster mass-to-light ratio we derive could not be so close to the
mean value of rich clusters (see Sect.~\ref{mlr}).

\subsection{The radio halo}
Here we consider the relation of the subcluster collision with
the radio halo of \mbox{1E0657-56}. We show in the left panel of
Fig.~\ref{rad_x} the 1.3 GHz radio image of the cluster, taken by
Liang et al. (\cite{lia00}) with ATCA, after subtraction of discrete
sources, and smoothing.  Radio contours are over-plotted on the
$R$-band image of the cluster obtained with FORS@VLT. The radio halo
seems to be centered onto the main cluster with an extension towards
the subcluster.

Cluster radio halos are quite uncommon. They are probably generated by
a population of ultra-relativistic electrons that emit synchrotron
radiation in the cluster magnetic field. The electrons could be
thermal in origin, belonging to the IC plasma, 
accelerated to high energies by energetic processes such as
cluster-cluster mergings. However, the merger event we are witnessing
might be too recent to explain the cluster radio halo. In fact,
according to the model of Brunetti et al. (\cite{bru01}), it takes
$\geq 0.6$ Gyr after the merger event to create a radio halo. On the
other hand, the radio halo of \mbox{1E0657-56} could have been powered
by other major subcluster collisions.  The presence of colliding
subclusters in the regions of massive clusters at $z \sim 0.3$ is
often seen in numerical simulations of hierarchical cosmologies
(e.g. Tormen \cite{tor98}), and the very elongated structure of
\mbox{1E0657-56} is certainly suggestive of accretion events along
large scale structure filaments.  The radio halo of \mbox{1E0657-56}
does have an extension towards the infalling subcluster, and this
feature is probably due to the displacement of the intra-cluster gas
by the recent collision event.

\subsection{The post-starburst population}
An additional consequence of a major merger could be the triggering of
a starburst activity in cluster galaxies (Bekki \cite{bek99}; Moss \&
Whittle \cite{mos00}; Girardi \& Biviano \cite{gb02} and references
therein), lasting $\approx 0.1$ Gyr.  We find no evidence for a
significant fraction of starburst galaxies in \mbox{1E0657-56}. On the
contrary, the cluster ELG fraction is rather low, compared to those of
other clusters at similar redshifts (e.g. the CNOC clusters, ELYC, and
the MORPHS clusters, D99), and it is similar to that of nearby
high-velocity dispersion clusters (see Fig.3 in Biviano et
al. \cite{biv97}). 

A significant number of cluster galaxies are instead found to be in a
post-starburst (PSB) phase, as indicated by their large
EW(H$\delta$). The fraction of PSB galaxies in \mbox{1E0657-56} ($\sim
20$\%) is similar to that of other clusters at the same redshift. The
PSB galaxies do not share the same kinematics of the main
\mbox{1E0657-56} galaxy population (see Sect.~\ref{morph}). This is an
indication for a recent (on a dynamical timescale) acceleration of
these galaxies. It is possible that both their kinematics and spectral
properties have been influenced by the subcluster collision, and that
we are observing the cluster just after the starburst phase has
ceased.  On the other hand, the PSB galaxies are not distributed along
the direction of the subcluster collision. Moreover, the mean velocity
of these galaxies (in the cluster rest frame) is $1076 \pm 235$
km~s$^{-1}$ lower than the mean velocity of the subcluster. It is then
unlikely that this galaxy population is related to the recent
collision event in \mbox{1E0657-56}. 

Alternatively, PSB galaxies could have evolved from recently
infallen late-type field galaxies.  In fact, PSB galaxies share the
projected distribution and kinematics of late-type galaxies (see
Fig.\ref{specspat} and Table~\ref{popvel}), suggestive of an
out-of-equilibrium dynamical state.  Significant ram-pressure
stripping occurs when galaxies cross the cluster core, sometime
preceded by an instantaneous burst of star formation (Fujita et
al. \cite{fuj01}; Vollmer et al.  \cite{vol01a}). Other episodes of
star formation can occur when the stripped galaxies emerging from the
core re-accrete part of the stripped gas on their disks (Vollmer et
al. \cite{vol00}, \cite{vol01b}). Should this scenario be correct,
the claimed r\^ole of subcluster collisions in the production of
starbursts should be reconsidered. A spectroscopic survey of the
outer regions of \mbox{1E0657-56} is needed in order to compare the
properties of galaxies in the cluster outskirts with those of galaxies
in the cluster core.

\section{Summary}
\label{summa}
We provide evidence for a recent ($\simeq 0.15$ Gyr) collision between
the very massive cluster \mbox{1E0657-56} and a low velocity
dispersion subcluster. A comparison between optical and X-ray
properties of the cluster suggests that a major collision has
occurred. The same comparison for the subcluster suggests that it
could be the remnant compact core of a moderately massive cluster.

Model timescales for the production of cluster radio halos tend to
exclude this collision event as the main responsible for the cluster
radio halo. On the other hand, the radio morphology of the cluster has
probably been distorted by this recent collision.

\mbox{1E0657-56} has a low fraction of star forming galaxies, rather
unexpected in current scenarios of starburst triggering by
cluster-cluster mergers. Furthermore, PSB galaxies are unlikely to be
related with the collision event, given their spatial and kinematical
distribution. We rather interpret them as field galaxies that have
recently fallen into the cluster gravitational potential.

\begin{acknowledgements}
We thank Dr. Haida Liang for kindly providing us with the radio image of the
cluster.  We acknowledge useful discussions with Marisa Girardi. 
We thank an anonymous referee for useful comments. This
research has made use of data obtained from the High Energy
Astrophysics Science Archive Research Center (HEASARC), provided by
NASA's Goddard Space Flight Center.
\end{acknowledgements}

\end{document}